\documentstyle[twocolumn,prl,aps]{revtex} 
 
\begin{document} 
\twocolumn[\hsize\textwidth\columnwidth\hsize\csname @twocolumnfalse\endcsname
\draft 
\preprint{} 
\title{A bosonic RVB description of doped antiferromagnet} 
\author{Z. Y. Weng, D. N. Sheng, and C. S. Ting} 
\address{Texas Center for Superconductivity and Department of Physics\\ 
University of Houston, Houston, TX 77204-5506 }  
\maketitle 
\date{today}
\begin{abstract} 
We propose a theory for doped antiferromagnet based on a {\it bosonic}
resonating-valence-bond (RVB) state with incorporating the phase string effect.
Both antiferromagnetic (AF) and superconducting phase transitions occur
naturally {\it within} such a bosonic RVB phase. Two 
distinct metallic regions -- underdoping and optimum-doping -- are also found to 
be a logic consequence, whose unique features explain the recent
neutron-scattering measurements in cuprates.

\end{abstract} 
 ]

It is well-known that bosonic RVB state\cite{liang,chen} has a {\it unrivaled 
high precision} in describing the $t-J$ model at no-hole case. A 
mean-field version\cite{aa} of such a bosonic spin description also provides 
a very convenient mathematical framework. Unfortunately, one soon falls into a 
stalemate once trying to get into metallic phase --- always 
encountering a spiral instability\cite{spiral,spiral1} at finite doping. 

Recently-revealed phase string effect\cite{string} has shed new light on this 
dilemma. When a doped hole moves on the spin background, the RVB spin pairs 
are broken up along the way and then re-paired, thanks to the
superexchange coupling. However, such a repairing can never be complete for a 
quantum spin-$1/2$ media. A phase string is
found\cite{string} to be left on the hole-path which is nonrepairable in 
low energy states. It indicates that a nontrivial Berry's phase 
will be acquired by a doped hole which slowly completes a closed-loop motion. 
Obviously, such a topological effect will be lost if the phase string effect is 
averaged out locally at each step of hopping, which suggests the spiral 
phase be an artifact of the inappropriate mean-field 
treatment.       

The phase string effect hidden in the bosonic RVB description implies the 
conventional slave-fermion formalism\cite{spiral1} of the $t-J$ model is
not a proper starting-point. A new {\it exact} 
reformulation\cite{string} of the model
has been obtained after using a canonical transformation to turn the
local phase string effect into an explicit nonlocal topological  
effect. Such a new representation is believed to be more suitable for a 
mean-field treatment in both 1D and 2D cases\cite{string}. In this
paper, we construct a generalized bosonic RVB mean-field theory based on this 
new formalism. Main results are summarized in the phase diagram shown in 
Fig. 1, where $\Delta^s\neq 0$ denotes the bosonic RVB phase which practically 
covers the whole experimentally interested temperature and doping regime. 
Two real phase transitions occur at low temperature {\it on} this bosonic RVB 
background, which are the insulating antiferromagnetic long range order 
(AFLRO) and superconducting condensation (SC). The former is due to a Bose 
condensation (BC) of bosonic spinons and the latter is due to a BC 
of bosonic holons. A striking feature is that the AFLRO cannot exist in the 
metallic phase but the spinon BC does persist into an underdoped metallic 
region, leading to a microscopic charge inhomogeneity. All of these, including 
the normal state, are controlled by the {\it single} bosonic RVB
order parameter $\Delta^s$, indicating the bosonic RVB pairing to
be a crucial driving force responsible for a whole spectrum of anomalies in the 
doped antiferromagnet. Some peculiar predications of the theory will be compared
with the experiments in cuprate superconductors.  
 
The $t-J$ model, $H_{t-J}=H_t + H_J$, may be generally expressed as
\begin{equation}\label{et}
H_t=-t\sum_{\langle ij\rangle}\hat{H}_{ij}
\hat{B}_{ji}+ H.c.,  
\end{equation}
\begin{equation}\label{ej}
H_J=-\frac J 2 \sum_{\langle ij\rangle} \hat{\Delta}^s_{ij}\left(\hat{\Delta}^s_{ij}\right)^{\dagger}.
\end{equation}
In the Schwinger-bonson, slave-fermion representation\cite{spiral1},
one has 
$\hat{H}_{ij}=f^{\dagger}_if_j$, $\hat{B}_{ji}=\sum_{\sigma}\sigma b^{\dagger}_{j\sigma}b_{i\sigma}$, and $\hat{\Delta}^s_{ij}=\sum_{\sigma}b_{i\sigma}b_{j-\sigma}$.
Here $f_i$ is a fermionic ``holon'' operator and $b_{i\sigma}$
is known as the Schwinger-boson operator. The bosonic RVB order parameter first introduced 
by Arovas and Auerbach\cite{aa} is defined as follows
\begin{equation}\label{ds}
\Delta^s=\langle \hat{\Delta}^s_{ij}\rangle.
\end{equation}
The sign $\sigma$ appearing in $\hat{B}_{ji}$ is the
source of phase string effect mentioned at the beginning. In
order to gain a finite hopping integral, i.e., $\langle \hat{B}_{ji}\rangle\neq 
0$, at the mean-field level, we see that up and down spins have to contribute 
differently to avoid cancellation, which always leads to a spiral twist. 
According to previous discussion\cite{string}, however, this mean-field 
procedure in doped case may be fundamentally flawed as the nontrivial 
topological 
effect of phase string is totally lost. To avoid this difficulty, by a canonical 
transformation\cite{string}, one may reformulate the model such that 
$\hat{B}_{ji}=\sum_{\sigma}e^{i\sigma A_{ji}^h}\bar{b}_{j\sigma}^{\dagger}
\bar{b}_{i \sigma}$, where the singular sign $\sigma$ is replaced by a link 
variable $e^{i\sigma A_{ji}^h}$. Correspondingly, $\hat{\Delta}^s_{ij}$ and 
$\hat{H}_{ij}$ are redefined by $\hat{\Delta}^s_{ij}=\sum_{\sigma}e^{-i\sigma
A_{ij}^h}\bar{b}_{i\sigma}\bar{b}_{j- \sigma}$ and $\hat{H}_{ij}=e^{iA_{ij}^f}h^{\dagger}_ih_j$, respectively. Here the nonlocal 
gauge field $A_{ij}^h$ is defined by a gauge-invariant condition 
$\sum_{C} A_{ij}^h=\pi N^h_C$ for an oriented closed-path $C$ with $N_C^h$ being the total hole number enclosed by $C$. Apparently $A_{ij}^h$ vanishes at 
zero-hole limit. And $A_{ij}^f$
satisfies $\sum_{C} A_{ij}^f=\pi \sum_{\sigma}\sigma
\sum_{l\in C} n_{l\sigma}^b-\Phi_C$
with $n_{l\sigma}^b =\bar{b}^{\dagger}_{\l\sigma}\bar{b}_{l\sigma}$
and $\Phi_C$ referring to a uniform flux with a strength of $\pi$-per-plaquette 
enclosed by $C$. Here new spinon and holon operators, $\bar{b}_{i\sigma}$ 
and $h_i$, are both {\it bosonic} as a peculiar result of the phase string 
effect.

Now the phase string effect is precisely tracked through the link variables,
$e^{i\sigma A_{ij}^h}$ and $e^{iA_{ij}^f}$, in $\hat{B}$, $\hat{\Delta}^s$,
and $\hat{H}$. We will still consider the mean-field solution characterized by
the RVB order parameter defined in Eq. (\ref{ds}). 
Similar to the no-hole case\cite{aa}, the spin part will be diagonalized by 
a Bogolubov transformation
\begin{equation}\label{bogo}
\bar{b}_{i\sigma}=\sum_m\left( u_m\gamma_{m\sigma}-v_{m}\gamma^{\dagger}_{m-\sigma}\right)e^{i\sigma \chi_m}\bar{w}_{m\sigma}(i).
\end{equation} 
Here $\gamma_{m\sigma}$ is an annihilation operator of spinon excitations and
the ``single-particle'' eigenfunction $\bar{w}_{m\sigma}(i)$ is
determined by  
\begin{equation}\label{ew}
\xi_m \bar{w}_{m\sigma}(i)=-J_s \sum_{j=nn(i)} e^{i\sigma A_{ij}^h}\bar{w}_{m\sigma}(j),
\end{equation}
with $J_s=J\Delta^s/2$. We explicitly introduce a phase factor 
$e^{i\sigma\chi_m}$ in Eq. (\ref{bogo}) 
to show a phase uncertainty in $\bar{w}_{m\sigma}$ which cannot be 
determined by Eq.(\ref{ew}) as it is a linear equation. It means that the
relation between $\bar{b}$ and $\gamma$ is not unique. Without changing
$\Delta^s$, such a ``phase'' freedom can only be fixed by optimizing the hopping 
integral $\langle \hat{B}_{ji}\rangle$. Suppose $\chi_m$ be a general function 
of the hole 
configuration. Then a maximum hopping $\langle \hat{B}_{ji}\rangle$ may be
achieved by a simple
phase shift $e^{i\sigma\chi_m}\rightarrow -sgn(\xi_m) e^{i\sigma\chi_m}$ each 
time a hole changes a site\cite{weng}. The coefficient $u_m$ and $v_m$ in 
Eq.(\ref{bogo}) 
are given by $(\lambda_m/E_m+1)^{1/2}/\sqrt{2}$ and $sgn(\xi_m)(\lambda_m/E_m-
1)^{1/2}/\sqrt{2}$, respectively. Here the spinon spectrum is
$E_m=\sqrt{\lambda_m^2-\xi_m^2}$, in which the hopping term only contributes to 
a shift to the Lagrangian multiplier $\lambda$ as $\lambda_m=\lambda-
J_h/J_s|\xi_m|$. Here the renormalized coupling constant 
$J_h= \langle{\hat{H}}\rangle t$ will be always chosen as $J_h=\delta J$ below
($\delta$ is the doping concentration).  
$\lambda$ is determined by the condition $\sum_i\langle b^{\dagger}_{i\sigma}b_{i\sigma}\rangle=N(1-\delta)$, or
\begin{equation}\label{lambda}
2-\delta =\frac 1 N \sum_m\frac{\lambda_m}{E_m}\coth \frac{\beta 
E_m}{2}+n_{BC}^b,
\end{equation}
where $\beta=1/k_BT$ and $n_{BC}^b$ represents the number of spinons per site 
staying at $E_m=0$ 
state if a Bose condensation of spinons occurs. In Fig. 1, the region of a 
nonzero $\Delta^s$ is shown which is self-consistently determined after
$A_{ij}^h$ in Eq.(\ref{ew}) is approximately replaced by a mean-field
$\bar{A}_{ij}^h$ defined by $\sum_C \bar{A}_{ij}^h=\pi\langle N_c^h\rangle$. The 
effect of fluctuations in $A_{ij}^h$ will be discussed below.

In the bosonic RVB description, $\Delta^s$ does not directly correspond to an
energy gap, in contrast to the fermionic RVB state\cite{rvb}. In fact, the 
spinon spectrum $E_m$ is known\cite{aa} to be {\it gapless}
at zero doping and zero temperature which ensures a BC of
spinons. In the new formulation, the transverse spin operator can be written 
as\cite{string}
\begin{equation}\label{s+}
S^+_i=\bar{b}^{\dagger}_{i\uparrow}\bar{b}_{i\downarrow}(-1)^ie^{i\Phi_i^h}
\end{equation}
in which 
\begin{equation}
\Phi_i^h=\sum_{l\neq i}\mbox{Im ln $(z_i-z_l)$} n_l^h,
\end{equation}
describes vortices (with vorticity $=1$) centered on holes 
($n^h_l=h^{\dagger}_ih_i$). In the absence of holes, a Bose condensation of
spinons will always give rise to $\langle S^+_i\rangle\propto (-1)^i$, i.e., an 
AFLRO. But in the presence of mobile holons --- in a metallic region
--- free vortices introduced by $\Phi^h_i$ will generally make $\langle 
S^+_i\rangle=0 $ even though spinons may be still Bose-condensed,
resembling a disordered phase in the 
Kosterlitz-Thouless transition. Only in an insulating phase where holes are
localized, the AFLRO may be still sustained as the vortex effect of $\Phi^h_i$ 
in Eq. (\ref{s+}) can be  ``screened'' through the compensation of a phase with 
opposite vorticities generated from spinons (After all, the phase string effect
is no longer effective if holes are localized). 

So the AFLRO should be absent in the metallic phase even though the spinon 
BC may persist in. One may then wonder what is the nature of such 
a metallic phase. In the following, we argue that a spinon BC phase in a 
metallic regime must be generally charge inhomogeneous. Recall that in the BC 
phase, $\lambda$ takes a value to make $E_m$ gapless 
such that $n_{BC}^b\neq 0$ can balance the difference between the left and right 
sides of Eq.(\ref{lambda}). Note that the $E_m=0$ state corresponds to the 
maximum of $|\xi_m|$, and thus it is related to those states at the band edge of
$\xi_m$ which is generally sensitive to the fluctuation of $A_{ij}^h$. As 
$A_{ij}^h$ is basically controlled by the holon density, the fluctuations of the
charge will then leads to a ``Lifshitz'' tail in $\xi_m$ and play an essential role in determining the $E_m=0$ state, which becomes a macroscopic state after the spinon Bose condensation. Such a state is generally associated with 
inhomogeneous 
hole configurations with the condensed spinons forming order in hole-deficient
region. Obviously, the 
detailed nature of the $E_m=0$ state will be sensitive to many factors, like the 
dynamics of holons which is beyond the present approximation. We will simply 
treat the fluctuation part $\delta A_{ij}^h=A_{ij}^h-
\bar{A}_{ij}^h$ in terms of a random flux description below. Those 
quantities like transition temperature $T_{BC}$ will not be our primary concern
here. Generally speaking, 
with the increase of doping, the reduction of the left-hand-side of 
Eq.(\ref{lambda}) will eventually make the BC contribution go away. In Fig.1
the shaded curve sketches such a BC region which basically defines an underdoped
metallic phase. 
(The dotted curve in Fig. 1 marks the insulating AFLRO phase in the dilute hole 
regime.)

On the other hand, the superconducting order parameter 
$\hat{\Delta}^{SC}_{ij}=\sum_{\sigma}
\sigma c_{i\sigma}c_{j-\sigma}$ can be expressed\cite{string} in the following 
form
\begin{equation}\label{sc}
\hat{\Delta}^{SC}_{ij}=\hat{\Delta}^s_{ij} \left(h_i^{\dagger}e^{\frac i 2
[\Phi^s_i-\phi^0_i]}\right)\left(h^{\dagger}_je^{\frac i 2 [\Phi^s_j-
\phi^0_j]}\right)(-1)^i,
\end{equation}
where 
\begin{equation}
\Phi_i^s=\sum_{l\neq i}\mbox{Im ln $(z_i-z_l)$}\sum_{\alpha}\alpha n_{l\alpha}^b
\end{equation}
describes vortices (anti-vortices) centered on up (down) spinons, and 
$\phi^0_i\rightarrow \phi_i^0\pm 2\pi$ after $i$ circles once along the loop of
a plaquette. In the 
present bosonic RVB description, spinons are always paired ($\Delta^s\neq 0$).
In order to have a superconducting condensation, then, bosonic holons have to 
undergo 
a Bose  condensation. The vortices described by $\Phi^s_i$ are all paired up in 
the ground state whose effect is minimal there. But at finite temperature, free 
vortices appear in $\Phi^s_i$ as spinons are thermally excited from the paired 
state. In order to achieve the 
phase coherence in Eq.(\ref{sc}), the condensed holons have to ``screen'' those
free vortices by forming supercurrents. A phase transition to normal state 
eventually happens when such ``screening'' fails which can be estimated as the 
free spinon number exceeds the holon number. The BC of holons will be
interrupted 
simultaneously. So the transition temperature $T_c$ may be determined by
\begin{equation}\label{tc}
\left.\frac{1}{N}\sum_{m}\frac{\lambda_m}{E_m}\sum_{\sigma}\langle 
\gamma_{m\sigma}^{\dagger}\gamma_{m\sigma}\rangle\right|_{T=T_c}=\kappa\delta,
\end{equation}
where $\kappa\sim 1$ and the left-hand-side represent the number of excited 
spinons determined from $\sum_{i\sigma} b^{\dagger}_{i\sigma}b_{i\sigma}$. 
$T_c$ calculated based on Eq.(\ref{tc}) is plotted in Fig. 1 as the dashed 
curve which is obtained under $A_{ij}^h\approx \bar{A}_{ij}^h$ in 
Eq.(\ref{ew}). This replacement should be considered to be an ``optimum'' case 
as the fluctuations of $A_{ij}^h$ generally reduces $T_c$. We will further 
discuss the optimum condition later. Finally, the symmetry of the SC order 
parameter may be determined as
\begin{equation}
\langle{\hat{\Delta}^s_{ii+\hat{x}}}\rangle/\langle{\hat{\Delta}^s_{ii+\hat{y}}}
\rangle =e^{-i\frac{1}{2}\sum_{\Box} \Delta \phi^0_{jk}}=-1,
\end{equation}
where $\sum_{\Box}$ denotes a summation of $\Delta \phi^0_{jk}\equiv \phi^0_j-\phi^0_k$ over four links of a plaquette and the result indicates a d-wave symmetry for the nearest-neighboring SC pairing.

Therefore, generally there are two temperature scales: $T_{BC}$ and $T_c$, in a
metallic phase. At low doping where $T_{BC}>T_c$, the charge inhomogeneity phase 
happens below $T_{BC}$ and further below $T_c$ holons are also expected to be
condensed into 
non-uniform regions in favor of the spin correlation energy. On the contrary, 
once $T_c> T_{BC}$, holons will experience Bose-condensation {\it first} and be 
{\it uniformly} distributed in real space since there is no pre-formed 
local spin 
ordering. Consequently, $A_{ij}^h$ may be replaced by $\bar{A}_{ij}^h$ with a substantial reduction of $\delta A_{ij}^h$ below $T_c$. In turn, the
spinon spectrum is qualitatively changed which prevents spinons from a
Bose condensing into an inhomogeneous phase at low-temperature (see below).
$T_c$ shown 
in Fig. 1 is estimated under such a condition and it optimizes $T_c$ as compared 
to the case with stronger fluctuations in $A_{ij}^h$. Beyond $T_c>T_{BC}$, a
crossover due to statistics transmutation may quickly set in as holons tend to 
be {\it always} Bose-condensed even at high temperature such that spinons have 
to be turned into fermions, which leads to the breakdown of the bosonic RVB 
state and is beyond the scope of the present paper.

How can two regions of metallic phase be distinguished by experiment?
The underdoped metallic region with a spinon BC must be a charge
inhomogeneity phase and also bears some resemblance to 
``pseudo-gap'' phenomenon. For example, both uniform spin susceptibility and 
transport resistivity will exhibit ``pseudo-gap'' behavior in this 
region. But we would like to focus on a more direct experimental signature here. The underdoped phase may be best characterized by a double-peak structure in low-energy region of local dynamic
spin susceptibility $\chi_L''(\omega)$ shown in Fig. 2. Here 
\begin{eqnarray}\label{chil}
& & \ \ \ \ \ \  \chi_L''(\omega)=\frac {\pi}{4}\sum_{mm'}K_{mm'} \left[1/2 (1+n(E_m)+n(E_m'))\right.\nonumber\\
& &\cdot(u_m^2v_{m'}^2+v_m^2u_{m'}^2)\delta(|\omega|-E_m-E_{m'})+(n(E_m) \nonumber\\
& &\left. -n(E_{m'}))(u_m^2u_{m'}^2+v^2_{m}v_{m'}^2)\delta(\omega+E_m-
E_{m'})\right],
\end{eqnarray}
where $\omega>0$ with $K_{mm'}\equiv \sum_{i\sigma}|w_{m\sigma}(i)|^2|w_{m'\sigma}(i)|^2$ and $n(E_m)$ as the Bose 
function. The lowest peak in Fig. 2 originates from 
excitations from the condensed spinons which can be explicitly sorted out from
Eq.(\ref{chil}) as
\begin{equation}\label{chilc}
\chi_c''(\omega)=\left(\frac{\pi}{4}n_{BC}^b\right)\sum_m K_{0m}\frac{\lambda_m}{E_m}\delta(\omega-E_m),
\end{equation} 
where $m=0$ refers to the BC state. This peak disappears above 
$T_{BC}$, while the second peak is contributed by usual spinon pairs excited 
from the vacuum. Here $\delta A_{ij}^h$ is treated as a
random flux, as noted before, with a strength of the flux per plaquette
$\delta\phi= 0.3\pi\delta$ at $\delta=1/7\approx 0.143$. Corresponding $T_{BC}$ 
is found to be $\sim 0.21J$.
In contrast, if the holon BC happens first such that $A^h_{ij}\approx 
\bar{A}_{ij}^h$, then the spinon BC is found absent below $T_c$ and only a 
single peak 
is left as shown in the insert of Fig. 2. Note that the sharpness of the peak is 
due to the Landau-level effect caused by $\bar{A}_{ij}^h$ in $E_m$ (for
simplicity we choose $\delta\phi=0$ here, i.e., $\delta 
A_{ij}^h$ is totally neglected below $T_c$). Other choices of parameters all 
yield similar two types of structure depending only on whether there is a 
spinon BC or not. 

Neutron-scattering measurement has indeed revealed a double-peak structure in 
$YBa_2Cu_3O_{6.5}$ compound recently\cite{bourges}, where the 
lower peak is located near $30$$meV$ and the second one is, around $60$$meV$, 
about twice bigger in energy as predicted by the theory (Fig. 2). Thus, this
underdoped material can be understood in the present theory as in the spinon BC 
phase. On the other hand, a ``resonance-like'' sharp peak at $41$$meV$ has been 
well-known for $YBa_2Cu_3O_7$ below $T_c$\cite{fong}, which is consistent with the case shown in the insert of Fig. 2 if $J\sim 100$$meV$. Namely, it 
corresponds 
to the uniform phase without the spinon BC, defined as an optimum metallic region in the theory with an optimized $T_c$. A single peak located at 
${\bf Q}_0=$($\pi$, $\pi$) is also identified in the momentum space at the 
``resonance'' energy in both the experiment and theory, suggesting an AF nature 
of spin fluctuations. But an incommensurate momentum structure is generally 
present at energy near the first peak shown in Fig. 2 in the spinon BC phase 
with charge inhomogeneity, which is sensitive to the nature of the BC
state. Finally, it is noted that in the bosonic RVB description, the 
double-layer coupling should not qualitatively change the above energy structure 
in the odd symmetry channel.

\acknowledgments
The present work is supported by a 
grant from the Robert A. Welch foundation, and 
by Texas Center for Superconductivity at University of Houston.\\

\figure{Fig. 1 The phase diagram of doped-antiferromagnet based the bosonic RVB 
description. The dotted and shaded curves sketch an insulating AFLRO phase and
an inhomogeneous metallic region, respectively, described by a spinon Bose 
condensation (BC). $SC$ indicates the superconducting condensation region 
determined under an optimal condition (see the text). The temperature $T$ is in units of $J$.}
\figure{Fig. 2 Local dynamic spin susceptibility $\chi_L''(\omega)$ vs. $\omega$ 
(in units of $J$) at $\delta=0.143$. Solid curve: $T=0$; ($\diamond$): $T=0.1$; 
($\times$): $T=0.2$; ($*$): $T=0.3$. Here $T_{BC}=0.21$ with $\delta\phi=0.3 \bar{\phi}$. The insert: $\chi_L''(\omega)$ vs. $\omega$ under the optimal 
condition: $\delta\phi=0$ at $T=0$. }

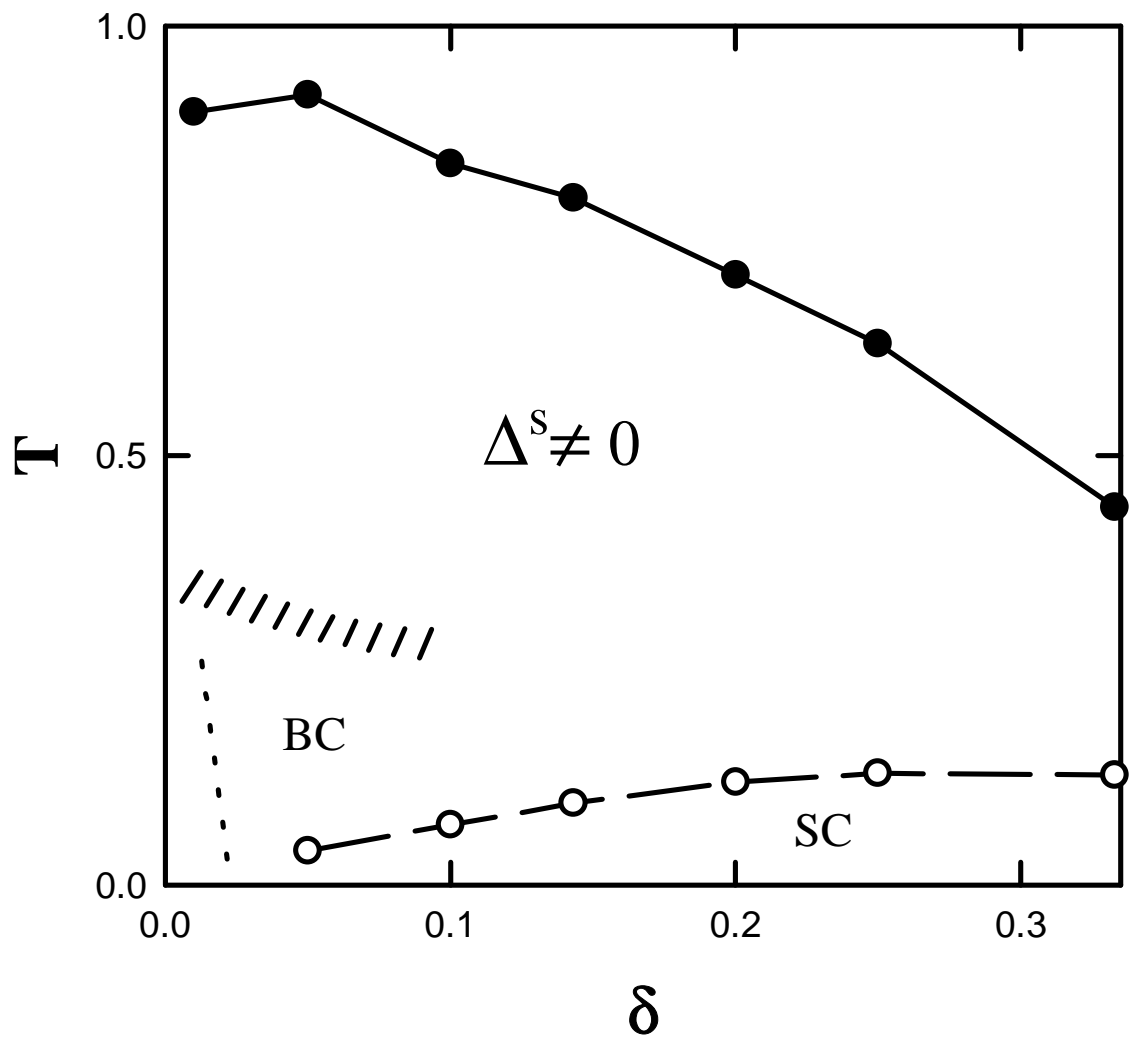
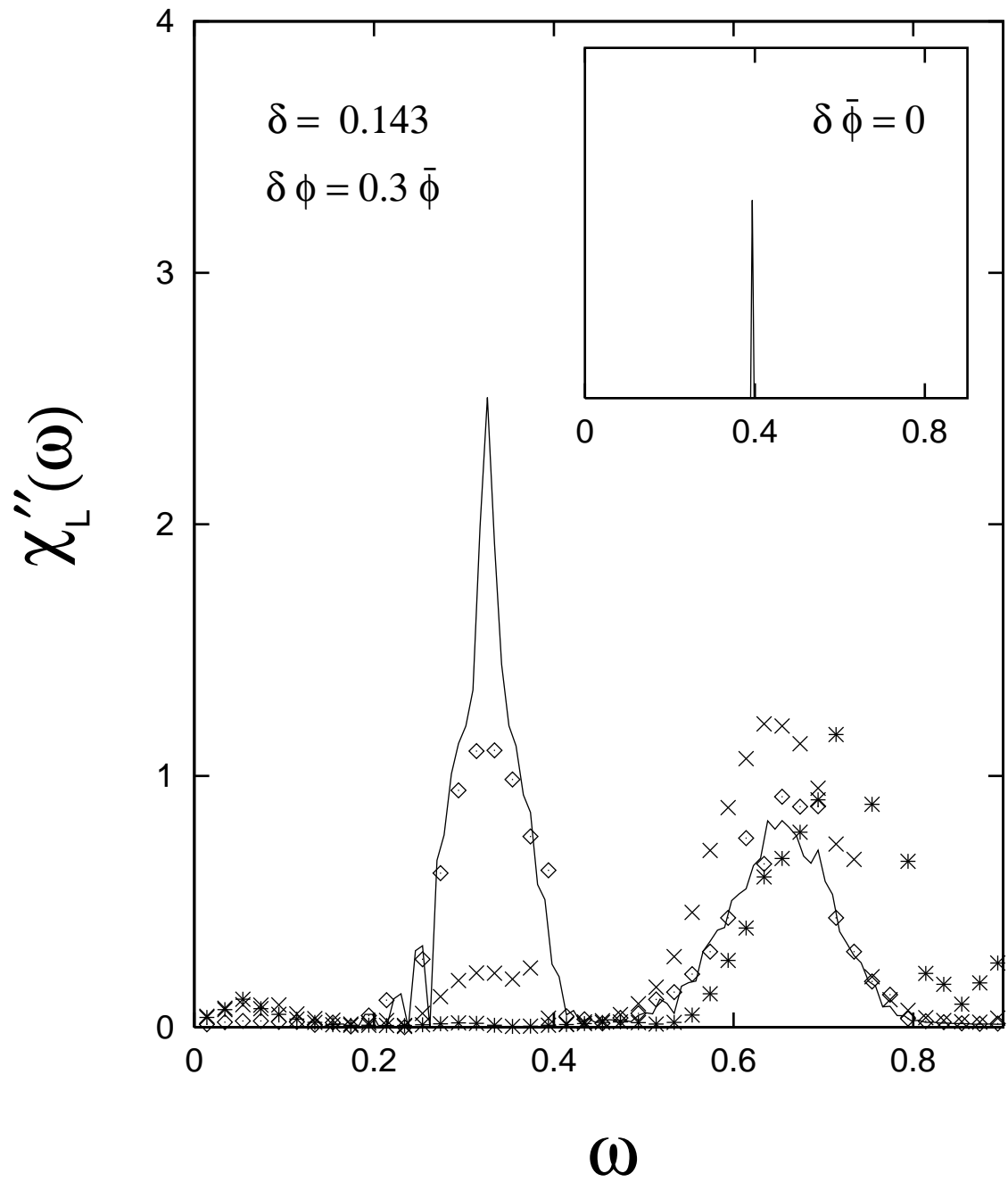
\end{document}